\documentclass[twocolumn,showpacs,preprintnumbers,amsmath,amssymb]{revtex4}

\usepackage{graphicx}
\usepackage{dcolumn}
\usepackage{bm}
\usepackage{color}
\usepackage{float}

\begin{document}


\title{Understanding fusion and its suppression for the $^{9}$Be projectile with different targets}
\author{V. Jha$^1$\footnote{vjha@barc.gov.in}}
\author{V. V. Parkar$^1$\footnote{vparkar@barc.gov.in}}
\author{S. Kailas$^{1,2}$\footnote{kailas@barc.gov.in}}

\affiliation{$^1$Nuclear Physics Division, Bhabha Atomic Research Centre, Mumbai - 400085, India}
\affiliation{$^2$UM-DAE Centre for Excellence in Basic Sciences, Mumbai - 400098, India}
\date{\today}

\begin{abstract}

The role of the breakup process and one neutron stripping on the near barrier fusion are investigated for the weakly bound projectile $^{9}$Be on  $^{28}$Si, $^{89}$Y, $^{124}$Sn, $^{144}$Sm and $^{208}$Pb targets.
Continuum-discretized coupled channels (CDCC) calculations for the breakup with a $^{8}$Be + n model of the $^{9}$Be nucleus and coupled reactions channels (CRC) calculations for the one neutron stripping to several single particle states in the target
are performed for these systems. A good description of the experimental fusion cross sections above the Coulomb barrier is obtained from the CDCC-CRC calculations for all the systems. The calculated  incomplete fusion probabilities for different target systems  are found to be consistent with the systematic behaviour of the complete fusion suppression factors as a function of target atomic mass, obtained from the experimental data.

\end{abstract}

\pacs{25.60.Pj, 25.70.Jj, 21.60.Gx, 24.10.Eq}
\maketitle

\section{\label{sec:Intro} Introduction}
Experiments using  stable weakly bound nuclei provide valuable avenues to understand the scattering and reaction with exotic radioactive nuclei. In this context, $^{6}$Li, $^{7}$Li and $^{9}$Be beams have been found extremely useful to probe the reaction dynamics around the Coulomb barrier \cite{Can,Keeley1}.
Unlike the reactions with tightly bound nuclei, a substantial contribution of breakup and transfer is observed in case of weakly bound nuclei. The breakup and transfer may be followed by subsequent fusion of the only a part of the projectile, a process called as incomplete fusion (ICF).
In this scenario, the fusion is classified in terms of complete fusion (CF) which refers to fusion of the whole projectile or all its fragments and the total fusion (TF), where the incomplete fusion (ICF) processes are also included.
In fusion measurements involving weakly bound nuclei, a large suppression of complete fusion cross section is observed at energies above the Coulomb barrier with respect to conventional coupled channels (CC) calculations \cite{Maha2,Maha4,Wu,Gas,Pra,Gomes1,Pal,rath,parkar,Harphool}, 
that either exclude the breakup and transfer couplings or include them only in an average way.
This suppression is commensurate with the measured ICF cross section \cite{Maha2}. Most of the measurements on suppression are performed with relatively heavy mass targets, because there is difficulty in separating contributions from
the CF and the ICF in case of light mass targets, as many of the evaporation residues coincide.

The theoretical calculations do not provide unequivocal answers about the coupling effects of breakup and transfer on the fusion cross sections and the CF suppression in case of weakly bound nuclei \cite{Hus,Tak,Das,Hag,Dia1,Dia2,Rus1,Keel,Ito}.
The observed experimental suppression of complete fusion is often explained by invoking two sets of principles. In the first approach, the coupled channel effects due to breakup and/or transfer are shown to alter the fusion cross section.
Continuum discretized coupled channels (CDCC) for breakup and Coupled reaction channels (CRC) calculations for transfer have been performed to study these dynamic effects arising due to coupling. The reduction in the fusion cross sections obtained due to the coupled channel effects of breakup process is found to be smaller
compared to the suppression seen from the data \cite{jha}. The continuum-continuum couplings have been responsible for most of the reduction in fusion cross sections with respect to the uncoupled calculations \cite{Dia1}. In some cases, the coupled channel approaches show a small enhancement at energies below the Coulomb barrier,  somewhat similar but less in magnitude than observed
in case of fusion with the stable nuclei \cite{Dia2}. Moreover, in the CDCC and CRC calculations, it is mostly the TF that is calculated, while the ICF and  CF calculations are not so straightforward, as explained in Ref.\ \cite{Thom1}. The second approach for explaining the CF suppression is based on simple empirical arguments,
where the CF cross sections can be estimated by subtracting the measured or calculated ICF from the TF cross sections \cite{Gomes}. The theoretical modeling for calculating the ICF component is still an open challenge, although a stochastic breakup model based on
classical trajectories to calculate the ICF has been developed in recent years \cite{Dia3}.

 The systematic behaviour of the fusion suppression factors and ICF probability as a function of target mass is not well understood, despite the CF experimental data being available for a number of projectile-target systems. Since, the amount of CF suppression observed in experiments with respect to CC calculations compares
well with the ICF cross section, the ICF probability ($P_{ICF}$), defined as $P_{ICF} = {\sigma_{ICF} / \sigma_{TF}}$, provides an indirect measure of CF suppression factors (F$_{CF}$). Recently, Gomes \textit{et al.} \cite{Gomes}, attempted to give a universal description of F$_{CF}$ for the CF of $^9$Be nucleus that is based on the estimation of the ICF yield derived from a
simple empirical relation given by Hinde \textit{et al.} \cite{Hinde}. The breakup probability function that depends on the gradient of the nuclear potential along with a exponential dependence on the surface separation was used to estimate the ICF probability. This simple model predicts that the ICF component and the F$_{CF}$
monotonically decrease with the target charge $Z_T$ for a given projectile. While the F$_{CF}$ extracted for the $^{9}$Be + $^{208}$Pb and $^{9}$Be + $^{144}$Sm are in qualitative agreement with this model, the CF suppression factors derived for the $^{9}$Be + $^{89}$Y and $^{9}$Be + $^{124}$Sn systems, both measured at BARC-TIFR pelletron, Mumbai are found not to follow this systematics.
This is in contrast to the results obtained in Ref.\ \cite{Raf}, where it is concluded that the CF suppression is nearly independent of target atomic number $Z_T$.

In this paper, a new method to calculate the ICF is employed that is based on absorption cross sections obtained from the CDCC calculations. Recently, it has been shown that a two body $^8$Be + n cluster structure of $^9$Be nucleus  provides a good description of the elastic scattering data for $^9$Be projectile with several target systems \cite{sanat,vivek13}. Here, the efficacy of this model is tested in explaining the fusion cross sections for the $^{9}$Be on different targets in a wide mass region ranging from the light to the heavy target.
In case of $^9$Be induced reactions, the $^8$Be + n breakup and the one neutron stripping processes are expected to have significant contribution. The calculations are performed to study the coupling effects of breakup and transfer channels on the fusion cross section using the CDCC-CRC approach.

\section{\label{sec:Caln} Calculation Details}
\begin{table*}[htbp]
\caption{\label{specfac} Energy levels of residual nuclei and spectroscopic amplitudes (SA) used in the CRC calculations.}\ \\
\begin{center}
\begin{tabular}{|ccc|ccc|ccc|ccc|ccc|}
\hline \multicolumn{3} {|c|} {$^{29}$Si}&
\multicolumn{3} {|c|} {$^{90}$Y}&
\multicolumn{3} {|c|} {$^{125}$Sn}&
\multicolumn{3} {|c|} {$^{145}$Sm}&
\multicolumn{3} {|c|} {$^{209}$Pb} \\ \hline E& J$^{\pi}$& SA & E& J$^{\pi}$& SA & E& J$^{\pi}$& SA & E& J$^{\pi}$& SA & E& J$^{\pi}$& SA
\\(MeV)&&&(MeV)&&&(MeV)&&&(MeV)&&&(MeV)&&\\ \hline
0.00& 1/2$^+$& 0.69 & 0.00& 2$^-$& 1.05& 0.00& 11/2$^-$& 0.52 & 0.00& 7/2$^-$& 0.78& 0.00& 9/2$^+$& 0.91 \\
1.27& 3/2$^+$& 0.83 & 0.20& 3$^-$& 1.03 & 0.03& 3/2$^+$& 0.55 & 0.89& 3/2$^-$& 0.66& 0.78& 11/2$^+$& 1.00 \\
2.03& 5/2$^+$& 0.22 & 1.22& 0$^-$& 1.05 & 0.22& 1/2$^+$& 0.52 & 1.11& 13/2$^+$& 0.81& 1.42& 15/2$^-$& 1.00 \\
3.62& 7/2$^-$& 0.62 & 1.38& 1$^-$& 1.05 & 1.26& 5/2$^+$& 0.17 & 1.43& 9/2$^-$& 0.92& 1.57& 5/2$^+$& 0.99 \\
4.93& 3/2$^-$& 0.84 & 1.96& 5$^+$& 0.52 & 1.36& 7/2$^+$& 0.17 & 1.61& 1/2$^-$& 0.91& 2.03& 1/2$^+$& 0.99 \\
5.95& 3/2$^+$& 0.37 & 2.25& 6$^+$& 0.54 & 1.54& 5/2$^+$& 0.02 & 1.66& 5/2$^-$& 0.64& 2.49& 7/2$^+$& 1.03 \\
6.38& 1/2$^-$& 0.78 & 2.84& 4$^-$& 0.67 & 2.75& 7/2$^-$& 0.15 & 1.79& 9/2$^-$& 0.58& 2.54& 3/2$^+$& 1.03 \\
& & & 2.94& 4$^-$& 0.67 & 3.41& 3/2$^-$& 0.42 & 2.71& 13/2$^+$& 0.55& & & \\
& & & 3.00& 5$^+$& 0.85 &4.01 & 1/2$^-$& 0.33 & & & & & & \\
& & & 3.05& 3$^-$& 0.67 & & & & & & & & & \\
& & & 4.07& 5$^+$& 0.88 & & & & & & & & & \\
\hline
\end{tabular}
\end{center}
\end{table*}

\begin{table}

\caption{\label{opt} Optical model potentials used in the calculations. For short range imaginary calculations, W$_{0}$ = 50.0 MeV, r$_{0}$ = 0.9 fm, a$_{0}$ = 0.25 fm was used. 
The radius  of the neutron potentials ($V_{n-T}$) is given as $R=r_o{A_T}^{1/3}$ while for the core-target potential ($V_{^8Be-T}$), 
the radius is given as $R=r_o({A_p}^{1/3}+{A_T}^{1/3})$ where $A_p$ and $A_T$ are the 
projectile and target mass numbers.}
\begin{center}
\begin{tabular}
{ccccc}
\hline
System & V$_{0}$ & r$_{0}$ & a$_{0}$ & Ref. \\
& (MeV) & (fm) & (fm) & \\
\hline \\

$^{8}$Be+$^{28}$Si & 39.69 & 1.16 & 0.59 &\cite{Win} \\
n+$^{28}$Si & 43.31 & 1.29 & 0.57 &\cite{Mori07} \\
$^{8}$Be+$^{89}$Y & 47.36 & 1.17 & 0.62 &\cite{Win} \\
n+$^{89}$Y & 43.10 & 1.28 & 0.57 &\cite{Mori07} \\
$^{8}$Be+$^{124}$Sn & 49.39 & 1.17 & 0.62 &\cite{Win} \\
n+$^{124}$Sn & 42.99 & 1.27 & 0.58 &\cite{Mori07} \\
$^{8}$Be+$^{144}$Sm & 50.28 & 1.18 & 0.63 &\cite{Win} \\
n+$^{144}$Sm & 42.93 & 1.26 & 0.58 &\cite{Mori07} \\
$^{8}$Be+$^{208}$Pb & 52.37 & 1.29 & 0.60 &\cite{sanat} \\
n+$^{208}$Pb & 42.75 & 1.24 & 0.61 &\cite{Mori07} \\
\hline
\end{tabular}
\end{center}
\end{table}

Fusion process for the $^9$Be + $^{28}$Si, $^{89}$Y, $^{124}$Sn, $^{144}$Sm and $^{208}$Pb target systems is studied using the coupled channels calculations with the CDCC-CRC approach. Calculations for the $^8$Be + n breakup of the $^9$Be and one neutron transfer to different single particle states in target have been performed for these systems.
Breakup calculations are performed with a three-body model for the projectile-target system using the CDCC method. The version FRXY.li of code FRESCO \cite{Thom88} is used for these calculations. A two-body $^8$Be + n cluster structure of $^9$Be  that has been shown to describe very nicely
the elastic scattering of $^9$Be on different target systems \cite{sanat,vivek13}, is used in the present calculations. Earlier this model has also been used for the calculations of fusion cross sections for the $^9$Be+$^{208}$Pb system \cite{sing2}.

The breakup of $^9$Be projectile is described by  the inelastic excitations of the $n$-$^{8}$Be ground state to the continuum that is induced by fragment-target potentials $V_{^8Be-T}$ and  $V_{n-T}$. The ground state wave-function of $^9$Be is generated using a potential with
Woods Saxon volume potential  and a spin-orbit component  taken from Ref.\ \cite{Lang77}. The ${1/2}^{+}$ and ${5/2}^{+}$ resonance states are generated  by using the same potential parameters as that of the ground state except for the potential depth,
which is varied to obtain the resonances with correct energies. The non-resonant continuum states are then generated with the same potential as that used for the resonance states. A relative angular momentum value of up to $l=4$ for the neutron core relative motion is taken for the calculations. The inclusion of higher $l$ values lead to less than $\approx$2 percent change in the calculated fusion cross section.
The continuum up to an energy $E_{max} \approx$ 7MeV above the $^8$Be + n breakup threshold is used in the calculations.

In addition to the breakup couplings, the effect of 1n-transfer couplings have also been investigated through a combined CDCC-CRC approach as explained in Ref.\ \cite{vivek13}, 
where the CDCC wave functions are used for the transfer calculations.
The neutron stripping channels; $^{28}$Si($^9$Be,$^8$Be)$^{29}$Si, $^{89}$Y($^9$Be,$^8$Be)$^{90}$Y, $^{124}$Sn($^9$Be,$^8$Be)$^{125}$Sn,  $^{144}$Sm($^9$Be,$^8$Be)$^{145}$Sm 
and $^{208}$Pb($^9$Be,$^8$Be)$^{209}$Pb have positive $Q$-values 6.81 MeV, 5.19 MeV, 4.07 MeV, 5.09 MeV and
2.27 MeV respectively. Several important excited states of the residual nucleus, as determined from the test calculations, are included in the final CRC calculations. These excited states used in the final calculations are given in Table\ \ref{specfac}.

The cluster folded potentials required in the CDCC calculations for constructing the $^9$Be + target interaction potential  are obtained using the two body core-target and valence-target potentials
$V_{^8Be-T}$ and  $V_{n-T}$ respectively. 
For the $V_{n-T}$, the neutron potentials are taken from Morillon et al. \cite {Mori07} for all the systems. For the 
core-target potential $V_{^8Be-T}$, the potentials obtained from Ref.~\cite{Win, sanat} are used. The final potential parameters used for all these systems are listed in Table\ \ref{opt}.

In the CDCC-CRC calculations, the fusion  cross sections can be obtained as the total absorption cross section, which is equal to the difference of the total reaction cross section $\sigma_R$ and the cross section of all explicitly
coupled direct reaction channels $\sigma_{D}$. The reaction cross sections are in turn obtained from the elastic scattering S-matrix elements, $S_l$ given by
\begin{equation}
\sigma_R = \sigma_{D} + \sigma_{abs} = \frac{\pi}{k^2}\sum_{l}(2l+1)(1-|S_l|^2)
\label{eq:1}
\end{equation}
Here, $\hbar k$ represents the relative momentum of the two nuclei in the entrance channel. CDCC-CRC calculations are performed with the  inclusion of
the imaginary part of the optical potential for the fragment-target interaction to account for the irreversible loss of flux from the coupled channels set. If all the dominant direct processes
such as breakup and transfer are included, the absorption cross section corresponds to the fusion cross section. A common approach is to include the short-range imaginary potentials to model the irreversible loss of flux.
The short-range imaginary potential ensures that the total flux  in the scattering channels decreases by the absorption when either one or both of the projectile fragments and the target nuclei are in the range of the potential inside the Coulomb barrier. 
The use of this short-range imaginary potential simulates the use of an incoming wave boundary condition
inside the Coulomb barrier. In the calculations for the weakly bound nuclei, the imaginary potentials can be included in  different ways and they correspond to different quantities that are calculated \cite{Thom1}.
The short-range imaginary potentials can be defined either in the coordinates of both projectile fragments relative to the target or in coordinates of only one of the projectile fragments relative to target .

In the calculations presented here, the fusion cross sections are first calculated by including the short-range imaginary potentials in the coordinates of both projectile fragments relative to the target. 
A Woods-Saxon potential with parameters W$_0$ = 50 MeV, r$_0$ = 0.9 fm, and a = 0.25 fm is used as the short range imaginary potential ($W_{SR}$) for 
the fragment-target optical potentials. The results depend very weakly on the geometry parameters of $W_{SR}$ for any larger depth. The inclusion of the imaginary potentials in the fragment-target coordinate system 
ensures the absorption of the c.m. of both $^8$Be and $n$ fragments in the fusion reaction and therefore, the calculated value
can be compared with the measured total fusion cross sections. 
The CDCC calculations are also performed where the $W_{SR}$ is present for only one of the projectile fragments relative to the target,
namely, either for the ${^8Be-T}$ part  or for the ${n-T}$ part.

The unambiguous calculation of ICF using the coupled channel is a complicated task as the absorption of flux in the coupled channels calculations includes varying contribution of the ICF component depending on the incident energy and the
system under study. As with the CF process, the ICF represents the flux that is lost irreversibly from the scattering channels, in this case due to absorption of only part of the projectile.
An approximate estimation of the ICF cross sections can be made using the absorption cross sections in the following way.
The CDCC calculations with the breakup couplings only are performed with three choices of optical potentials, where $W_{SR}$ is used for i) both the
projectile fragments relative to the target ($PotA$) ii)  the ${^8Be-T}$ part only ($PotB$) and iii) the ${n-T}$ part only ($PotC$). 
 However, in all these calculations, an additional $W_{SR}$ without any real part is also present 
in the center of mass of the whole projectile for the projectile-target radial motion \cite{Dia1}. The use of additional imaginary potential is justified as
the results are found to be insensitive to any larger depth of the $W_{SR}$ than what is used. The inclusion of the extra $W_{SR}$ is necessary only in the case, where $PotC$ 
is used, to correctly calculate the CF component corresponding to ${^9Be}$ + target part. In the other two cases, the additional $W_{SR}$ does not matter, as the results are 
nearly the same whether it is included or not in the calculation. The absorption cross sections in three cases represent cross sections for i) complete fusion ($\sigma_{CF}$)  + $^8$Be partial fusion ($\sigma_{ICF_{^8Be}}$) + $n$ 
partial fusion ($\sigma_{ICF_n}$) ii) $\sigma_{CF}$ + $\sigma_{ICF_{^8Be}}$  and iii) $\sigma_{CF}$ + $\sigma_{ICF_n}$, respectively. These three calculations together are used to estimate the $\sigma_{ICF_{^8Be}}$ and 
$\sigma_{ICF_n}$ explicitly. In fact, $\sigma_{ICF_{^8Be}}$ is equivalent to $\sigma_{ICF_\alpha}$, 
as it is more likely that only one of the $\alpha$ particles may fuse with the target nucleus, since the $\alpha$ particles from the decay of 
$^8$Be nucleus are emitted back to back in the $^8$Be rest 
frame. The total ICF can be then evaluated as sum of $\sigma_{ICF_\alpha}$ and $\sigma_{ICF_n}$.

In addition, ICF can also arise from the 1n transfer into bound states of the target nucleus followed by the breakup of $^8$Be nucleus and the subsequent absorption of 
only one of the $\alpha$ particles. In fact, the transfer process has been found to predominantly trigger the breakup process in sub-barrier reactions of other weakly bound projectiles  $^{6,7}$Li 
on different targets \cite{Luo13}. The combined process of transfer followed by breakup and absorption of one $\alpha$ particle can be considered as an important source of ICF formation. The transfer cross sections calculated using CDCC-CRC calculations with $PotA$ for all the targets are taken as an approximate measure of the ICF formation 
due to transfer followed by breakup. Here, a reasonable assumption is made that 1n transfer is always followed by the breakup of the $^8$Be nucleus into two $\alpha$ 
particles and the capture of one of the $\alpha$ particle. It must be mentioned that the breakup and transfer calculations performed in this work, only take into account the $^8$Be+n breakup of $^9$Be or 1n transfer to the $^8$Be$_{g.s.}$. 
The breakup and ICF is possible also through the $^5$He+$^4$He two body decay \cite{Hen}, and the transfer via the $^8$Be$_{2^+}$ state. However, these components are estimated to be small compared to the dominant modes considered here \cite{sanat} and hence, these have not been considered.
\section{\label{sec:Result} Results and Discussion}\subsection{Fusion cross-sections}
\begin{figure}[!ht]
\includegraphics[scale=0.79,trim=0.0cm 5.0cm 13cm 1.0cm,clip=true]{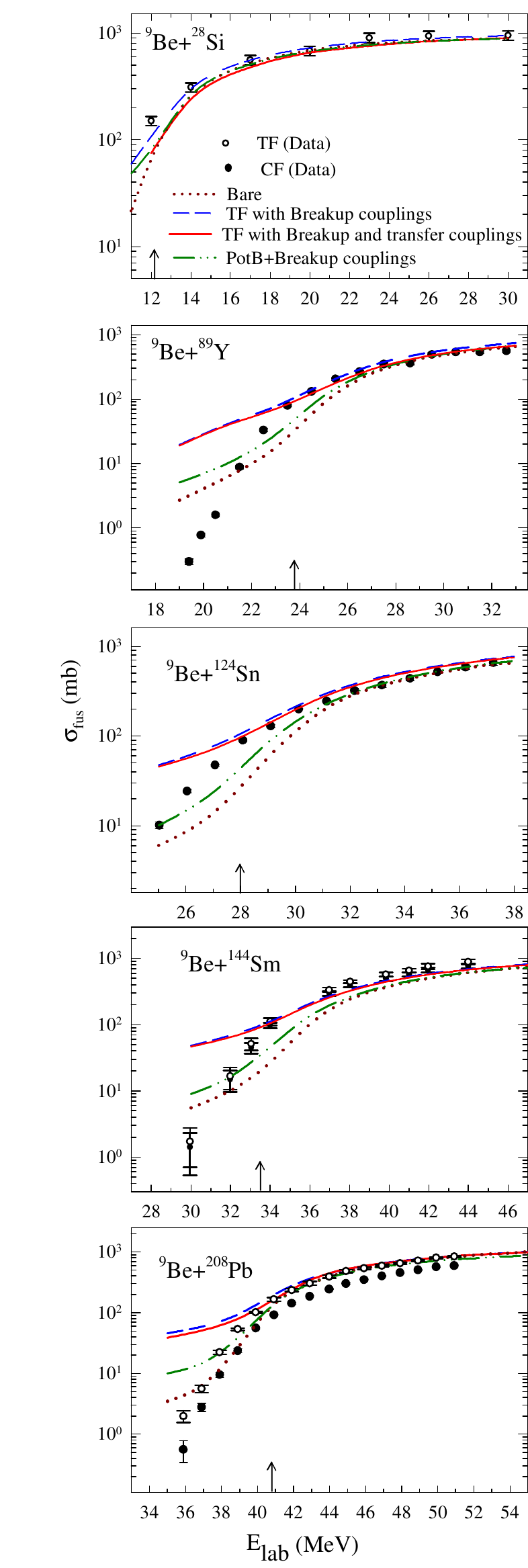}
\caption{\label{fig1} (Color online) Measured complete fusion (CF) and total fusion (TF) derived as CF + ICF (incomplete fusion) cross sections for $^9$Be + $^{28}$Si \cite{Bodek}, $^{89}$Y \cite{Pal}, $^{124}$Sn \cite{parkar}, $^{144}$Sm \cite{Gomes1} and $^{208}$Pb \cite{Maha4} target systems are compared
to calculated cross sections. The calculations with only bare potential, by inclusion of breakup couplings and by inclusion of breakup plus transfer couplings are shown. The calculations  performed using
potential with $PotB$ by inclusion of breakup couplings only are also shown. The arrows indicate the position of Coulomb barriers for each system.}
\end{figure}

The fusion excitation functions are calculated for various target systems to enable a direct comparison with the data.
The  fusion data for various systems along with the results of calculations are shown in Fig.\ \ref{fig1}. The absorption cross sections at different energies obtained from the CDCC-CRC calculations using $PotA$ optical potentials are compared with the measured fusion cross sections taken from the literature \cite{Bodek,Pal,parkar,Gomes1,Maha4}. 
In addition to the experimental total fusion cross sections, the complete fusion data is also plotted, wherever it is available.
The calculations with only the bare potentials are shown by dotted line, whereas the calculations that include the breakup couplings and the breakup + transfer couplings are shown by dashed  lines and the solid lines respectively. 
In comparison to the calculation with the bare potential, the calculations with the breakup couplings show a general enhancement of the fusion cross section for all target systems for the whole energy range. 
The enhancement is very pronounced at energies below the Coulomb barrier for all target systems, 
except for the $^9$Be+$^{28}$Si case, where data below the barrier is not available.
This observation of enhancement may be linked to the attractive real and absorptive imaginary polarization potentials that are obtained due to the $E1$ 
couplings within the $^8$Be+n model of $^9$Be breakup \cite{vivek13}. Inclusion of single-neutron stripping coupling does not give any significant effects, except a small reduction of the fusion cross section over the whole energy range. Again, this reduction seems to be consistent with the repulsive polarization potentials that arise due to the coupling of $+Q$-value transfer channels. 
However, the transfer coupling effects on fusion are comparatively smaller than the large effect observed in the near-barrier elastic scattering of $^9$Be+$^{208}$Pb system \cite{sanat}. 
The light system $^9$Be+$^{28}$Si has also significant transfer coupling effect despite having the higher Q-value for the 1n transfer reaction. 

A reasonably good description of the fusion data for all the target systems is obtained by the CDCC-CRC calculations using $PotA$ optical potentials
as given by the solid lines. Particularly, the available total fusion data for $^9$Be + $^{144}$Sm and $^9$Be + $^{208}$Pb systems are well described by the calculations, 
at above barrier energies. This is the energy region from where the suppression of the complete fusion is evaluated in terms of reduction with respect to coupled channels calculations.
The discrepancy observed between the calculation and the measured data at lower energies indicates that a weaker absorption of flux is required from the breakup and transfer channels into the fusion process than that predicted by the calculations.
Indeed, the CDCC calculations with the potential where the imaginary potential for the n-target system is switched off ($PotB$),
give a better description of the data  at the energies below the Coulomb barrier. The calculations using $PotB$ with only breakup couplings
are shown by dashed-dot-dot line in Fig.\ \ref{fig1}. Similar calculations using the core-target potential only have also been performed by Ito \textit{et al.} \cite{Ito},
for the fusion reactions of neutron-halo nuclei with a three-body time-dependent wave-packet method. The calculated cross section in this case, corresponds 
to the complete fusion and the incomplete core fusion, where the neutron escapes.
\subsection{Breakup, transfer probabilities and ICF}
\begin{figure}[!ht]
\centering
\includegraphics[scale=0.75, trim=0.0cm 3cm 10cm 0.0cm,clip=true]{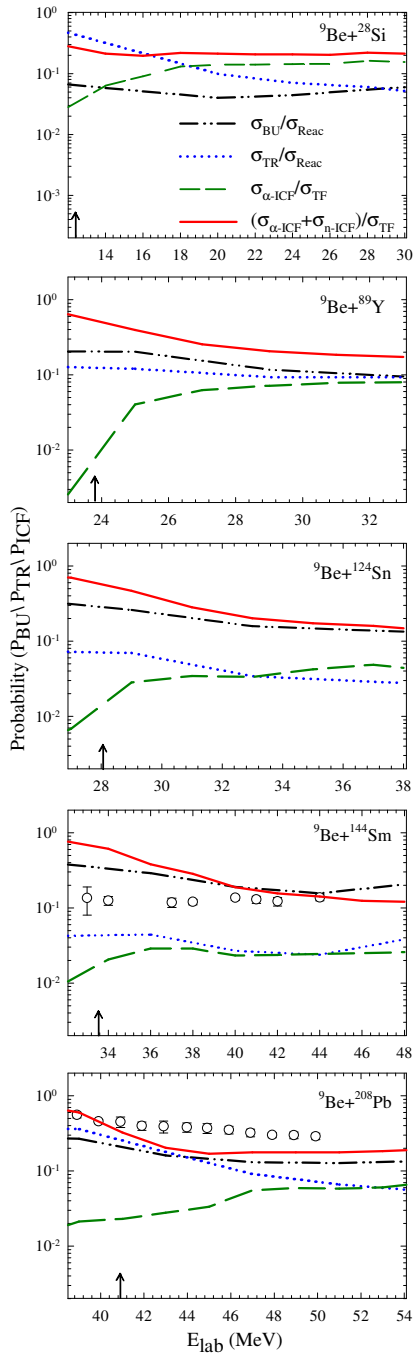}
\caption{\label{btr} (Color online) The calculated breakup probabilities ($P_{BU}$) and transfer probabilities ($P_{TR}$) obtained as the ratio of breakup and transfer cross sections and
the reaction cross sections are shown for a range of energies. The probabilities of the incomplete fusion due to breakup process for the neutron only ($P_{ICF_n}$) and combined 
ICF probability for both neutron and $\alpha$ ($P_{ICF_n}$ + $P_{ICF_\alpha}$) are shown. The Experimental ICF probabilities available for two systems $^9$Be+$^{144}$Sm \cite{Gomes1} and $^{208}$Pb \cite{Maha4} are also plotted
and compared with the  calculated $P_{ICF_n}$ + $P_{ICF_\alpha}$ values.}
\end{figure}

The CDCC-CRC calculations using the potentials  having short range imaginary potentials for both projectile fragments,
are utilized to investigate the systematic behaviour of  breakup and transfer cross sections as a function of incident energy.
The breakup process is expected to be the dominant process of  $\alpha$ production for the heavier systems. In contrast, the transfer process will depend on the structure of the target and the final residual nuclei.
The breakup probabilities ($P_{BU}$) and transfer probabilities ($P_{TR}$) are calculated
as the ratio of calculated breakup and transfer cross sections with the calculated reaction cross sections at each energy using CDCC-CRC calculations
with $PotA$ optical potentials.
The plots of $P_{BU}$ and $P_{TR}$ are shown by dashed-dot-dot and dotted lines respectively, in Fig.\ \ref{btr} for different target systems.
The breakup probabilities  remain approximately constant over the energy range above the Coulomb barrier while there is a small increase
below the barrier for all target systems, except for the $^9$Be + $^{28}$Si system.
The increase at the  sub-barrier energies, is consistent with the measured breakup probabilities in Ref.\cite{Raf} which shows an exponential rise with increasing energy.
At energies above the barrier, the probability of charged fragment capture by the target is approximately constant leading to a constant removal of flux from the breakup or transfer channel at these energies.
 The breakup and transfer contributions at higher energies are comparable for the $^{28}$Si and $^{89}$Y cases, while the
 breakup dominates over the transfer contribution in other cases.
At lower energies, the transfer contribution is significant only in the $^{28}$Si and $^{208}$Pb cases.
It is quite interesting to note that there is an almost constant variation of the  $P_{BU}$ at higher energies 
for the systems with different nuclear sizes and structure.

As pointed out earlier, the unambiguous calculation of ICF is not so straightforward.
In the present case, the $^8$Be + n  breakup of $^9$Be and one neutron transfer from $^9$Be are expected to be the dominant sources of ICF formation and inclusive $\alpha$ production.
Therefore, an accurate calculation of ICF cross section would require to follow and calculate the absorption probability
of the $\alpha$ particles and neutron after the breakup and transfer. As described before, neutron incomplete fusion ($\sigma_{ICF_n}$) and the $\alpha$ incomplete fusion
 ($\sigma_{ICF_\alpha}$) due to breakup process only, can be calculated using the absorption cross sections from the different choices of imaginary potential using CDCC calculations. The probabilities for the ICF of neutron ($P_{ICF_n}$) 
 and $\alpha$ ($P_{ICF_\alpha}$) are calculated as the ratio of calculated $\sigma_{ICF_n}$ and $\sigma_{ICF_\alpha}$ with the total fusion cross sections calculated using CDCC calculations 
 with  $PotA$ at each energy. The contribution of $P_{ICF_n}$ and $P_{ICF_n}$ + $P_{ICF_\alpha}$ for all the systems are shown in Fig.\ \ref{btr} by the dashed lines 
 and solid lines, respectively. These plots show that at lower energies $P_{ICF_n}$ is  dominant compared to the $P_{ICF_\alpha}$.

The calculated ICF fractions as described above, can also be used to estimate the inclusive $\alpha$ cross section.
The $\alpha$ production results  from the ICF as the other fragments $n$ and $\alpha$ from these processes may fuse with the target or the residual nuclei. In addition, the $\alpha$ production takes place also from
the exclusive breakup and transfer processes, where the resulting $^8$Be decays into two $\alpha$ particles.
Therefore, using the breakup and transfer cross sections along with the ICF cross sections, a measure of inclusive $\alpha$ production cross section can be obtained.

The ICF probabilities ($P_{ICF}$) extracted from the experimental data for two systems $^9$Be + $^{144}$Sm and $^9$Be + $^{208}$Pb are also plotted in Fig.\ \ref{btr}. 
It is compared with the calculated values of $P_{ICF_n}$ + $P_{ICF_\alpha}$ for these systems.
 The comparison is qualitative in view of the fact that the experimental measurements of ICF  do not distinguish whether the neutron is absorbed or not.
\subsection{ICF and Fusion Suppression}
\begin{figure}
\includegraphics[width=90mm,height=75mm]{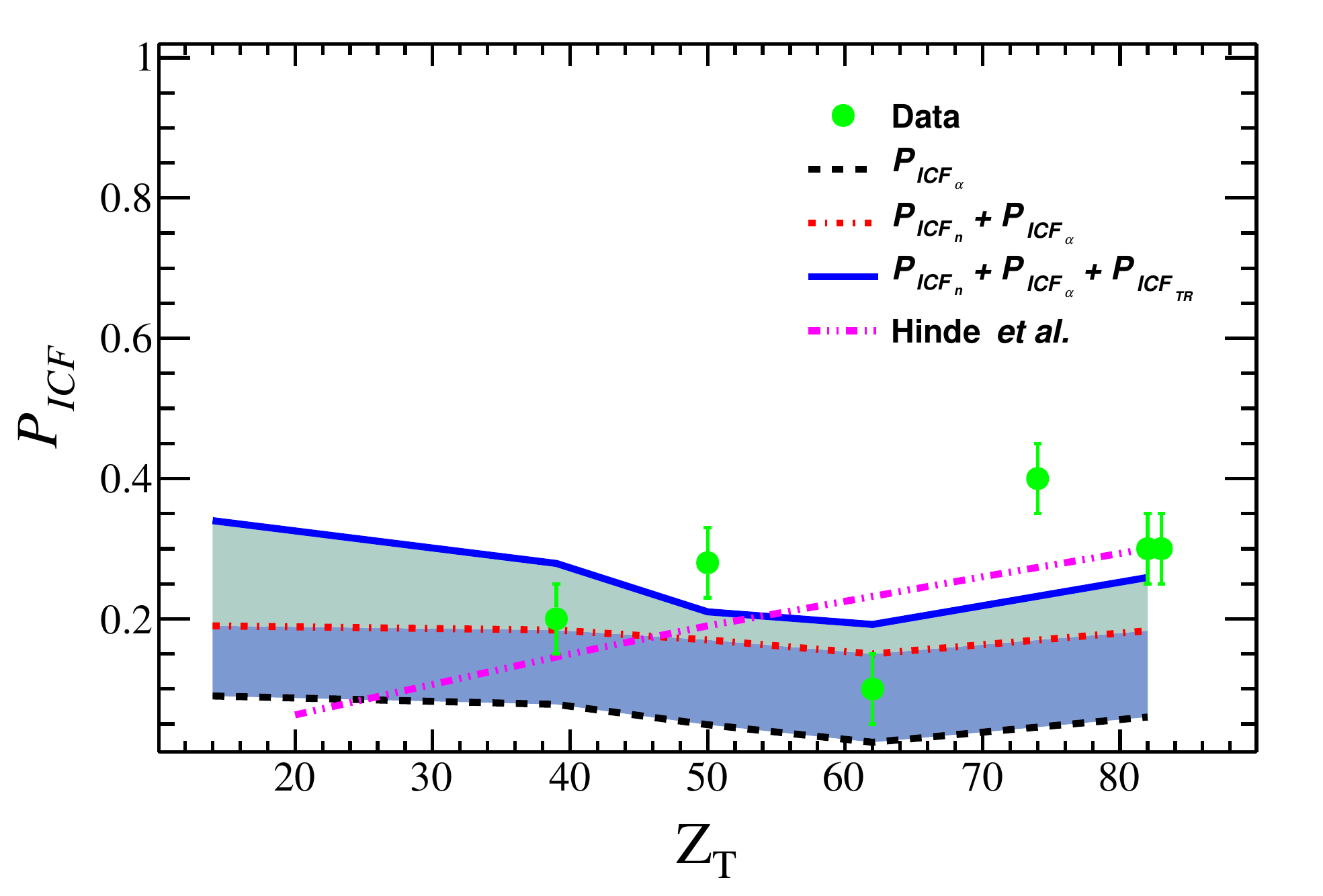}
\caption{\label{bup_com} (Color online) The calculated ICF probabilities ($P_{ICF}$) are compared with the suppression factors derived from  the
experimental data \cite{Pal,parkar,Gomes1,Fang,Maha4,Maha2010}. The region between values of $P_{ICF_\alpha}$ and $P_{ICF_n}$ + $P_{ICF_\alpha}$  for different targets are shown as the lower shaded region.
The behaviour of ICF contribution due to transfer is included by adding $P_{ICF_{TR}}$ to $P_{ICF_n}$ + $P_{ICF_\alpha}$ and the summed quantity 
is shown by the upper shaded region. All quantities are calculated at $1.3 V_b$. For comparison, the calculations using the model described in Ref.\ \cite{Hinde}
as given in Ref.\ \cite{Gomes} are also plotted.}
\end{figure}

The ICF probability ($P_{ICF}$) and the CF suppression factors ($F_{CF}$) are related quantities as the measured ICF cross sections are found to have the same magnitude as the difference of TF and CF cross sections \cite{Maha2}.
In the case of $^9$Be fusion, the measurements of the CF doesn't necessarily exclude the ICF component due to neutron fusion. The measured ICF may be only due to the fusion of $\alpha$ resulting from $^8$Be decay,
subsequent to the $^8$Be + n breakup. Therefore, the quantities  $P_{ICF_\alpha}$ and $P_{ICF_n}$ + $P_{ICF_\alpha}$ can be taken to represent the lower and upper limits  of measured $P_{ICF}$.
These quantities evaluated at the lab energy value $1.3 V_b$,  is plotted as shaded region between the dashed and dash - dotted line in Fig.\ \ref{bup_com}, where $V_b$ is the Coulomb barrier for the respective system.
For comparison, the fusion suppression factors $F_{CF}$ \cite{Pal,parkar,Gomes1,Fang,Maha4,Maha2010} extracted from the measured data for all the systems are also shown in Fig.\ \ref{bup_com}. 
The value of $P_{ICF}$ obtained from the present calculation are consistent with the values extracted from the
experimental data for the $^9$Be + $^{89}$Y and $^9$Be + $^{144}$Sm systems. 
 
The contribution of 1n transfer process to the ICF have been ignored till now. Experimentally, it is not always possible to distinguish the ICF from the transfer process, and the measured 
ICF usually includes the contribution of the  transfer processes \cite{Fang}. An approximate estimate of this contribution can be obtained 
by a quantity $P_{ICF_{TR}}$, which is defined as the ratio of transfer cross section with the total fusion cross sections that are calculated using 
the CDCC-CRC calculations with $PotA$. It must be remarked though, while the variation of ICF probability ($P_{ICF}$) due to breakup process 
can be taken as a continuous behaviour with the target mass, the $P_{ICF_{TR}}$ only indicates the behaviour for the systems for which the calculations are performed. 
The region between $P_{ICF_n}$ + $P_{ICF_\alpha}$ and the 
$P_{ICF_n}$ + $P_{ICF_\alpha}$ + $P_{ICF_{TR}}$ at the lab energy value $1.3 V_b$ is shown by the upper shaded region between dash-dotted line and solid line.
With this modified prescription of including the 1n transfer contributions,  a reasonable explanation of all fusion suppession data are obtained 
for which the calculations are performed. The large fusion suppression observed in $^9$Be + $^{184}$W fusion data  may be due to even higher transfer contribution
as the measured stripping cross section in that case suggests \cite{Fang}.
Interestingly, the data of $^9$Be+$^{89}$Y and $^{124}$Sn are found  consistent with the calculations contrary to the viewpoint expressed
in Ref.~\cite{Gomes}.  For comparison, the empirical prediction by Hinde \textit{et al.} \cite{Hinde} based on geometrical assumptions that predicts $P_{ICF}$ decreases with
target charge, due to the relatively smaller importance of the Coulomb breakup, is also plotted in Fig.\ \ref{bup_com} by the dot-dot-dashed line. 
The significant value of the $P_{ICF}$ predicted by the present calculations for the light system $^9$Be + $^{28}$Si and its nearly constant behaviour for all target masses, 
is at variance with the empirical prediction of Hinde \textit{et al.} \cite{Hinde}. 
The reasonable agreement of the data with the calculations without any normalization for wide range of target systems validate the estimation of 
ICF by the method employed here.

The nearly constant behaviour of the $P_{ICF}$ observed here, is in agreement with the calculated behaviour given in Ref.~\cite{Raf} for the $P_{ICF}$ of the targets in high $Z_T$ range. The small variation observed in the $P_{ICF}$ can be ascribed to the optical potentials employed in the calculations that describe the elastic scattering and fusion data simultaneously.
It seems that the CF suppression is a universal behaviour for the $^9$Be projectile with target mass ranging from heavy to light targets. Similar behaviour
is also observed in the experimental data of complete fusion with the other weakly bound nucleus $^6$Li \cite{Harphool}.
This behavior of $^9$Be CF suppression is also consistent with the direct measurement of above-barrier ICF of $^{6,7}$Li and $^{10}$B, incident on a range of heavy targets,
for which the systematics extracted in Ref.~\cite{Gas} showed their CF suppression factor to be independent of $Z_P Z_T$ within their experimental uncertainties.
The present results show that the phenomena of complete fusion suppression at above-barrier energies being almost independent of the target mass may be a feature with all weakly bound nuclei.
\section{\label{sec:Sum} Summary}
In summary, we have studied the effect of breakup and 1n-transfer couplings  on the fusion of $^9$Be  with $^{28}$Si, $^{89}$Y, $^{124}$Sn, $^{144}$Sm and $^{208}$Pb targets,
that spans a large target mass region, from light to heavy. The experimental fusion cross section data available around the Coulomb barrier have been utilized for these investigations.
The CDCC-CRC calculations have been performed by including the one neutron stripping and  $^8$Be + n breakup for $^9$Be to study their relative importance.
A good description of the above barrier data is obtained for all the systems that have been considered.
In general, an enhancement in the fusion cross section is obtained due to the coupling effects of breakup channels.
The 1n transfer channels are found to give a small reduction in the fusion cross section values with respect to uncoupled values.
This behaviour is supported by the nature of polarization potentials derived for some of these  systems in our earlier work \cite{vivek13}.
The significant contributions of 1n-transfer are obtained for the $^9$Be + $^{208}$Pb and  $^9$Be + $^{28}$Si systems, specially at the lower energies.

The  breakup and transfer probabilities calculated by CDCC-CRC calculations show a constant variation at energies above the barrier.
The calculated absorption cross sections from the CDCC calculations using three choices of interior imaginary potential are further utilized to obtain a measure of ICF cross section.
The ICF probabilities due to breakup at higher energies obtained from the absorption cross sections for the neutron and $\alpha$ partial fusions are in good agreement with the behaviour of measured complete fusion suppression factors. The CF suppression factors calculated as the ICF probabilities show a systematic behaviour with respect to different target masses and they remain approximately constant at energies above the barrier for all the systems considered. The ICF contribution due to transfer varies depending on the structure of target and the residual nuclei.
The exclusive measurements of ICF and CF cross sections with different weakly bound projectiles, especially in the light target mass region, wherever possible,
are needed to further verify this proposition. Simultaneously, an integrated theoretical modeling of the ICF and CF processes will be helpful
to understand the phenomena of complete fusion suppression in case of weakly bound projectiles.

\begin{acknowledgments}
We thank Mr. S. K. Pandit and Dr. H. Kumawat for useful discussions regarding the work. One of the authors (V.V.P) acknowledges the financial support through the INSPIRE Faculty program, from Department of Science and Technology, Govt. of India in carrying out these investigations.
\end{acknowledgments}


\begin{thebibliography}{99}
\bibitem{Can}  L. F. Canto, P.R.S. Gomes, R. Donangelo, and M. S. Hussein, Phys. Rep. {\bf 424}, 1 (2006) and references therein.
\bibitem{Keeley1}  N. Keeley, R. Raabe, N. Alamanos, and J. L. Sida, Prog. Part. Nucl. Phys. {\bf 59}, 579 (2007) and references therein.
\bibitem{Maha2} M. Dasgupta, D. J. Hinde, R. D. Butt, R. M. Anjos, A. C. Berriman, N. Carlin, P. R. S. Gomes, C. R. Morton, J. O. Newton, A. Szanto de Toledo, and K. Hagino, Phys. Rev. Lett. {\bf 82}, 1395 (1999).
\bibitem{Maha4} M. Dasgupta, P. R. S. Gomes, D. J. Hinde, S. B. Moraes, R. M. Anjos, A. C. Berriman, R. D. Butt, N. Carlin, J. Lubian, C. R. Morton, J. O. Newton, and A. Szanto de Toledo, Phys. Rev. {\bf C 70}, 024606 (2004)
\bibitem{Gas} L. R. Gasques, D. J. Hinde, M. Dasgupta, A. Mukherjee, and R. G. Thomas, Phys. Rev. {\bf C 79}, 034605 (2009).
\bibitem{Pal} C. S. Palshetkar, S. Santra, A. Chatterjee, K. Ramachandran, S. Thakur, S. K. Pandit, K. Mahata, A. Shrivastava, V. V. Parkar, and V. Nanal, Phys. Rev. {\bf C 82}, 044608 (2010).
\bibitem{parkar} V. V. Parkar, R. Palit, Sushil K. Sharma, B. S. Naidu, S. Santra, P. K. Joshi, P. K. Rath, K. Mahata, K. Ramachandran, T. Trivedi, and A. Raghav, Phys. Rev. {\bf C 82}, 054601 (2010).
\bibitem{Gomes1} P. R. S. Gomes, I. Padron, E. Crema, O. A. Capurro, J. O. Fernandez Niello, A. Arazi, G. V. Marti, J. Lubian, M. Trotta, A. J. Pacheco, J. E. Testoni, M. D. Rodriguez, M. E. Ortega, L. C. Chamon, R. M. Anjos, R. Veiga, M. Dasgupta, D. J. Hinde, and K. Hagino, Phys. Rev. {\bf C 73}, 064606 (2006).
\bibitem{Wu} Y. W. Wu, Z. H. Liu, C. J. Li, H. Q. Zhang, M. Ruan, F. Yang, Z. C. Li, M. Trotta, and K. Hagino, Phys. Rev. {\bf C 68}, 044605 (2003).
\bibitem{rath} P. K. Rath, S. Santra, N. L. Singh, R. Tripathi, V. V. Parkar, B. K. Nayak, K. Mahata, R. Palit, Suresh Kumar, S. Mukherjee, S. Appannababu, and R. K. Choudhury, Phys. Rev. C {\bf 79}, 051601(R) (2009).
\bibitem{Pra} M. K. Pradhan, A. Mukherjee, P. Basu, A. Goswami, R. Kshetri, S. Roy, P. Roy Chowdhury, M. Saha-Sarkar, R. Palit, V. V. Parkar, S. Santra, and M. Ray, Phys. Rev. {\bf C 83}, 064606 (2011).
\bibitem{Harphool} H. Kumawat, V. Jha, V. V. Parkar, B. J. Roy, S. K. Pandit, R. Palit, P. K. Rath, C. S. Palshetkar, Sushil K. Sharma, Shital Thakur, A. K. Mohanty, A. Chatterjee, and S. Kailas, Phys. Rev. C {\bf 86}, 024607 (2012).
\bibitem{Hus} M. S. Hussein, M. P. Pato, L. F. Canto, and R. Donangelo, Phys. Rev. {\bf C 46}, 377 (1992).
\bibitem{Tak} N. Takigawa, M. Kuratani, and H. Sagawa, Phys. Rev. {\bf C 47}, R2470 (1993).
\bibitem{Das} C. H. Dasso and A. Vitturi, Phys. Rev. {\bf C 50}, R12 (1994).
\bibitem{Hag} K. Hagino, A. Vitturi, C. H. Dasso, and S. M. Lenzi,  Phys. Rev. {\bf C 61}, 037602 (2000).
\bibitem{Dia1} A. Diaz-Torres and I. J. Thompson Phys. Rev. {\bf C 65}, 024606 (2002).
\bibitem{Dia2} A. Diaz-Torres, I. J. Thompson, and C. Beck, Phys. Rev. {\bf C 68}, 044607 (2003).
\bibitem{Rus1} K. Rusek, N. alamanos, N. Keeley, V. Lapoux, and A. Pakou, Phys. Rev. {\bf C 70}, 014603 (2004).
\bibitem{Keel} N. Keeley and N. Alamanos, Phys. Rev. {\bf C 77}, 054602 (2008).
\bibitem{Ito} M. Ito, K. Yabana, T. Nakatsukasa, and M. Ueda, Phys. Lett. {\bf B 637}, 53 (2006).
\bibitem{jha} V. Jha and S. Kailas, Phys. Rev. {\bf C 80}, 034607 (2009).
\bibitem{Thom1} I. J. Thompson and A. Diaz-Torres, Prog. Theor. Phys. Supplement {\bf 154}, 69 (2004).
\bibitem{Gomes} P. R. S. Gomes, R. Linares, J. Lubian, C. C. Lopes, E. N. Cardozo, B. H. F. Pereira, and I. Padron, Phys. Rev. {\bf C 84}, 014615 (2011)
\bibitem{Dia3} A. Diaz-Torres, D. J. Hinde, J. A. Tostevin, M. Dasgupta, and L. R. Gasques, Phys. Rev. Lett. {\bf 98}, 152701 (2007).
\bibitem{Hinde} D. J. Hinde, M. Dasgupta, B. R. Fulton, C. R. Morton, R. J. Wooliscroft, A. C. Berriman, and K. Hagino, Phys. Rev. Lett. {\bf 89}, 272701 (2002).
\bibitem{Raf} R. Rafiei, R. du Rietz, D. H. Luong, D. J. Hinde, M. Dasgupta, M. Evers, and A. Diaz-Torres, Phys. Rev. {\bf C 81}, 024601 (2010)
\bibitem{sanat} S. K. Pandit, V. Jha, K. Mahata, S. Santra, C. S. Palshetkar, K. Ramachandran, V. V. Parkar, A. Shrivastava, H. Kumawat, B. J. Roy, A.Chatterjee, and S.Kailas, Phys. Rev. {\bf C 84}, 031601(R) (2011).
\bibitem{vivek13} V. V. Parkar, V. Jha, S. K. Pandit,  S. Santra, and S. Kailas, Phys. Rev. {\bf C 87}, 034602 (2013).
\bibitem{Thom88} I. J. Thompson, Comput. Phys. Rep. {\bf 7}, 167 (1988).
\bibitem{sing2} C. Signorini, Eur. Phys. J. {\bf A 13}, 129 (2002).
\bibitem{Lang77} J. Lang, R. M$\ddot{\textrm{u}}$ller, J. Untern$\ddot{\textrm{a}}$hrer, L. Jarczyk, B. Kamys, and A. Strzalkowski, Phys. Rev. {\bf C 16}, 1448 (1977).
\bibitem{Bal77} R. Balzer, M. Hugi, B. Kamys, J. Lang, R. M$\ddot{\textrm{u}}$ller, E. Unugricht, J. Untern$\ddot{\textrm{a}}$hre, L. Jarczyk, and A. Strzalkowski, Nucl. Phys. {\bf A 293}, 518 (1977).
\bibitem{Mori07} B. Morillon and P. Romain, Phys. Rev. {\bf C 76}, 044601 (2007).
\bibitem{Win} R. A. Broglia and A. Winther, Heavy Ion Reactions, Lecture Notes Vol. I (Addison-Wesley, Redwood City, CA, 1991), p. 114.
\bibitem{Luo13}  D. H. Luong, M. Dasgupta, D. J. Hinde, R. du Rietz, R. Rafiei, C. J. Lin, M. Evers and A. Diaz-Torres, Phys. Rev. {\bf C 88}, 034613 (2013).
\bibitem{Hen} H. Esbensen, Phys. Rev. {\bf C 81}, 034606 (2010).
\bibitem{Bodek} K. Bodek, M. Hugi, J. Lang, R. M$\ddot{\textrm{u}}$ller, E. Ungricht, K. Jankowski, W. Zipper, L. Jarczyk, A. Strzalkowski, G. Willim, and H. Witala, Nucl. Phys. {\bf A 339}, 353 (1980).
\bibitem{Fang} Y. D. Fang, P. R. S. Gomes, J. Lubian, X. H. Zhou, Y. H. Zhang, J. L. Han, M. L. Liu, Y. Zheng, S. Guo, J. G. Wang, Y. H. Qiang, Z. G. Wang, X. G. Wu, C. Y. He, Y. Zheng, C. B. Li, S. P. Hu, and S. H. Yao, Phys. Rev. {\bf C 87}, 024604 (2013).
\bibitem{Maha2010} M. Dasgupta, D. J. Hinde, S. L. Sheehy, and B. Bouriquet, Phys. Rev. {\bf C 81}, 024608 (2010).

\end{thebibliography}
\end{document}